\documentstyle[epsfig,eqsecnum,aps]{revtex}
\begin{document}

\title{Fission widths of hot nuclei from Langevin dynamics}

\author{ Gargi Chaudhuri \thanks{E-mail:gargi@veccal.ernet.in} and
Santanu Pal\thanks{E-mail:santanu@veccal.ernet.in}}
\address{Variable  Energy  Cyclotron  Centre,  1/AF Bidhan Nagar,
Calcutta 700 064, India}
%\date{\today }
\maketitle
\begin{abstract}
Fission dynamics of
excited nuclei is studied in the framework of Langevin equation.
The one body wall-and-window friction is used as the
dissipative force in the Langevin equation.
In addition to the usual wall formula friction, the chaos
weighted wall formula developed earlier to
account for nonintegrability of single-particle motion within the nuclear
volume is also considered here. The fission rate calculated with the chaos
weighted
wall formula is found to be faster by about a factor of two than that
obtained with the usual wall friction. The systematic dependence of
fission width on temperature and spin of the fissioning nucleus is investigated
and a simple parametric form of fission width is obtained.
\end{abstract}

\section{INTRODUCTION}

Fission of highly excited compound nuclei formed in heavy ion induced fusion
reactions has emerged as a topic of considerable theoretical and experimental
interest in recent years. Multiplicity measurements of light
particles and photons emitted in the prescission stage
 strongly suggest \cite{thoe} that fission is
a much slower process for hot nuclei than that determined from the statistical
model of Bohr and Wheeler \cite{bw} based on phase space arguements. This led
to a revival of theoretical studies based on the original work of Kramers
\cite{kram} who considered fission of excited nuclei as a consequence of
thermal fluctuations. Dynamical models for fission
based on Fokker-Planck equation
\cite{wm1,wm2}
and Langevin equation \cite{abe,fro1} were subsequently developed.

The most extensive application of Langevin equation to study fission dynamics
was made by Froebrich {\it et. al} \cite{fro1}. A combined dynamical and
statistical model for fission was employed in their calculations where a
switching over to a statistical model description was made when the fission
process reached the stationary regime. Fission widths which are required for
the statistical branch of the calculation  were obtained from
the stationary limits
of fission rates from Langevin equation \cite{fro2}.
The dissipative property of nuclei is an important input to
the Langevin
dynamical calculations, the choice of which is not yet fully settled and
continues to be an open question.
A detailed
comparison \cite{fro3}
of the calculated fission probability and prescission neutron multiplicity
excitation functions for a number of nuclei with
the experimental data led to a phenomenological shape dependent
nuclear friction.  The phenomenological friction turned out to be
considerably smaller ($\sim 10\%$) than the standard wall formula value for
nuclear friction
for compact
shapes of the fissioning nucleus whereas a strong increase of this friction
was found to be necessary  at large deformations. A clear physical picture
for such a
friction is yet to be developed and the present work is an effort in this
direction.

The wall formula for nuclear friction was developed by Blocki {\it et al.}
\cite{blocki} in a simple classical picture of one-body dissipation.
It was also derived from a formal theory based on classical
linear response theory \cite{koo}. One crucial assumption of the wall formula
concerns the randomization of the particle (nucleon) motion due to
the successive collisions it suffers at the nuclear surface. In other words,
a complete mixing in the classical phase space of the particle motion is
required.
It was early realized \cite{blocki,koo} that any deviation from this
randomization
assumption would give rise to a reduced strength of the wall formula.
Further, Nix and Sierk suggested \cite{nix1,nix2} in their analysis
of mean fragment
kinetic energy data that the dissipation is about four times weaker than that
predicted by the wall plus window formula of one-body dissipation.
However, it is only recently that
a modification of the wall formula has been proposed \cite{pal1} in
which the full randomization assumption is relaxed in order to make it
applicable to systems in which the mixing in phase space is partial.
Considering only those chaotic particle trajectories which arise
due to irregularity
in the shape of the one-body potential and  which are responsible for
irreversible energy tranfer, a modified friction coefficient was
obtained in Ref.\cite{pal1}. In what follows, we shall use the
term "chaos weighted
wall formula" (CWWF) for this modified friction in order to distinguish
it from the original wall formula (WF) friction. As was shown
in Ref.\cite{pal1},
the CWWF friction coefficient
${\eta_{cwwf}}$ will be given as,

\begin{eqnarray}
\label{fric}
\eta_{cwwf} = \mu \eta_{wf},
\label{seqn1}
\end{eqnarray}

\noindent where ${\eta_{wf}}$ is the friction coefficient as was given
by the original wall
formula \cite{blocki} and ${\mu}$ is a measure of chaos (chaoticity) in the
single-particle
 motion and depends on the instantaneous shape of the nucleus. The
value of chaoticity $\mu$ changes from 0 to 1 as the nucleus evolves from
the spherical shape to a highly deformed one. The CWWF friction is thus
much smaller than WF friction for compact nuclear shapes while they become
closer at large deformations.
The CWWF friction was subsequently found \cite{pal2,pal3} to describe
satisfactorily the
collective energy damping of cavities containing classical particles and
undergoing time-dependent shape evolutions.
Thus the supression of the strength of
wall formula friction achieved in chaos weighted
wall formula suggests that chaos in single particle motion (rather lack of it)
can provide a physical explanation for the reduction in strength of friction
for compact nuclear shapes  as required in the
phenomenological friction of Ref.\cite{fro3} and this has
motivated us to
apply CWWF friction to fission dynamics in the present work.

We shall present in this paper a systematic study of fission rates by using
both CWWF and WF frictions in Langevin equation. The aim of our study is
twofold. First, we would like to find the effect of introducing the
chaoticity factor in friction on fission rate at different excitation
energies and spins of the compound nucleus. The second one concerns a
parametric representation of the fission widths the need for which arises
as follows. Fission width is an essential input along with particle and
$\gamma$ widths for a statistical theory in the stationary branch of compound
nucleus decay. Kramers \cite{kram}  obtained a simple expression for the
stationary
fission width assuming a large separation between the
saddle and scission points and a constant friction. Gontchar {\it et. al}
\cite{fro2}
later derived a more general expression taking the scission point explicitly
into account but still assuming a constant friction coefficient. The CWWF
friction however is not constant and is strongly shape dependent and hence the
corresponding stationary fission width cannot be analytically obtained.
Thus it becomes necessary to find a suitable parametrization of the
numerically obtained stationary fission widths using CWWF friction in order
to use them in the statistical regime of the compound nucleus decay. We shall
concentrate upon the parametric representation of fission widths in the
present work while the application of CWWF friction in a full dynamical
plus statistical model will be reported in a future publication.

We shall describe the Langevin equation along with the necessary inputs
as used in the present calculation in the next section. The calculated
fission rates and the systematics of the stationary fission widths will be
given in Sec.III. A summary of the results will be presented in the last
section.

\section{LANGEVIN EQUATION FOR FISSION}
\subsection { Nuclear shape, potential and inertia}

In order to specify the collective coordinates for a dynamical description
of nuclear fission, we will use the $c,h,\alpha$ parametrisation of
Brack {\it et. al} \cite{brack}. We will consider only symmetric fission
($\alpha=0$)
and will further neglect the neck degree of freedom ($h=0$) to simplify the
calculation. The surface of a nucleus of mass number $A$ will then
be defined as,
\begin{eqnarray}
\rho^2(z) &=& \left(1 - \frac{z^2}{c_o^2}\right)(a_oc_o^2 + b_oz^2),
\label{seqn2}
\end{eqnarray}
where
\begin{eqnarray}
c_o &=& cR, \nonumber\\
 R &=& 1.16 A^{1 \over 3}, \nonumber
\end{eqnarray}
and
\begin{eqnarray}
a_o &=& \frac{1}{c^3} - \frac{b_o}{5},\nonumber\\
b_o &=& \frac{c-1}{2} ,\nonumber
\end{eqnarray}
in cylindrical coordinates for the elongation parameter $c$. Considering
$c$ and its conjugate momentum $p$ as the dynamical variables, the Langevin
equation in one dimension will be given \cite{ab2} as,
\begin{eqnarray}
\frac{dp}{dt} &=& -\frac{p^2}{2} \frac{\partial}{\partial c}\left({1 \over m}
\right) -
   \frac{\partial F}{\partial c} - \eta \dot c + R(t), \nonumber\\
\frac{dc}{dt} &=& \frac{p}{m} .
\label{seqn3}
\end{eqnarray}

\noindent In the above equations, $m$ and $\eta$ are the shape-dependent
collective
inertia and friction coefficients respectively. The free energy of the system
is denoted by $F$ while $R(t)$ represents the random part of
the interaction
between the fission degree of freedom and the rest of the nuclear degrees of
freedom considered as a thermal bath in the present picture.

We will use the Werner-Wheeler approximation for incompressible irrotational
flow to calculate the collective inertia \cite{davies}. The driving force
in a thermodynamic system should be derived from the free energy
for which we will use the following expression valid
for the Fermi gas model \cite{fro3},
\begin{equation}
F(c,T) = V(c) - a(c) T^2,
\label{seqn4}
\end{equation}

\noindent where $T$ is the temperature of the system and
$a(c)$ is the coordinate
dependent level density parameter which is given as \cite{balian},
\begin{equation}
a(c) = a_vA + a_sA^{2 \over 3} B_s(c).
\label{seqn5}
\end{equation}

\noindent The values for the parameters $a_{v}$, $a_{s}$ and the
dimensionless surface
area $B_{s}$ are chosen following Ref.\cite{fro3}.

For the potential energy $V(c)$, it is only the deformation dependent part of
it which is relevant for our calculation. This deformation dependent potential
energy is obtained from the finite range liquid drop model \cite{sierk}
where we calculate the generalized
nuclear energy by double folding
the uniform density
within the surface (Eq.\ref{seqn2}) with a
Yukawa-plus-exponential potential. The Coulomb energy is obtained by double
folding another
Yukawa function with the density distribution. The various input parameters
are taken from Ref.\cite{sierk} where they were determined from fitting
fission barriers of a wide range of nuclei. The centrifugal part of the
potential is calculated using the rigid body moment of inertia.

The instantaneous random force $R(t)$ plays a very crucial role in the
Langevin description of nuclear fission. As a result of receiving incessant
random kicks, the
fission degree of freedom can finally pick up enough kinetic energy to overcome
the fission barrier. This random force is modelled after that of a typical
Brownian motion and is assumed to have a stochastic nature with a Gaussian
distribution whose
average is zero \cite{abe}. It is further assumed that $R(t)$ has extremely
short correlation time implying that the intrinsic nuclear dynamics is
Markovian. Consequently the strength of the random force  can be obtained
from the
fluctuation-dissipation theorem and the properties of $R(t)$ can be
written as,
\begin{eqnarray}
\langle R(t) \rangle &=& 0 ,\nonumber\\
\langle R(t)R(t^{\prime})\rangle &=& 2\eta T \delta(t-t^{\prime}) .
\label{seqn6}
\end{eqnarray}

\subsection {One-body dissipation}

One-body dissipation was used more successfully in fission dynamics than
two-body viscosity in
the past \cite{abe,ab2}. Accordingly,
we shall consider the one-body wall-and-window dissipation \cite{blocki}
to account for the
friction coefficient $\eta$ in the Langevin equation. For the one-body wall
dissipation, we shall use the chaos weighted wall formula (Eq.\ref{seqn1})
introduced
in the preceeding section. In order to arrive at this expression, the particle
trajectories moving in the one-body nuclear potential were identified as
either regular or chaotic depending on their nature of time evolution
\cite{pal1,pal2}. Originating from a given point near the nuclear surface and
moving in a given direction, a regular trajectory closes smoothly in phase
space. On the other hand, another trajectory leaving the same point but in a
different direction could be a chaotic one which does not close in phase
space. Considering the contributions of these two types of trajectories
separately, it was argued in Refs. \cite{pal1,pal2}  that only the chaotic
trajectories give rise to
the irreversible energy transfer and the resulting friction coefficient
acting on the wall motion will be as given in Eq.\ref{seqn1} . The
chaoticity $\mu$ is a measure of chaos in the single-particle motion
of the nucleons within the nuclear volume and in the present classical picture
will be given as the average fraction of the trajectories which are chaotic
when the sampling is done uniformly over the nuclear surface.

The chaoticity is a specific property of the nonintegrability of the nuclear
shape. Thus it is required to be calculated for all possible shapes upto
the scission configuration. A typical calculation of chaoticity for a given
shape proceeds as follows. The initial coordinates of a classical trajectory
starting from the nuclear surface is chosen by sampling of a suitably defined
set of random numbers such that all the initial coordinates follow an uniform
distribution over the nuclear surface. The initial direction of the trajectory
is also chosen randomly and its  Lyapunov exponent  is then
obtained by following the trajectory for a considerable length of time.
Each trajectory is identified as regular or chaotic by considering the
magnitude of its Lyapunov exponent and the nature of its variation with time.
The details of this procedure are given in Ref.\cite{blocki2}.

\begin{figure}[htb]
\centering{\
\epsfig{figure=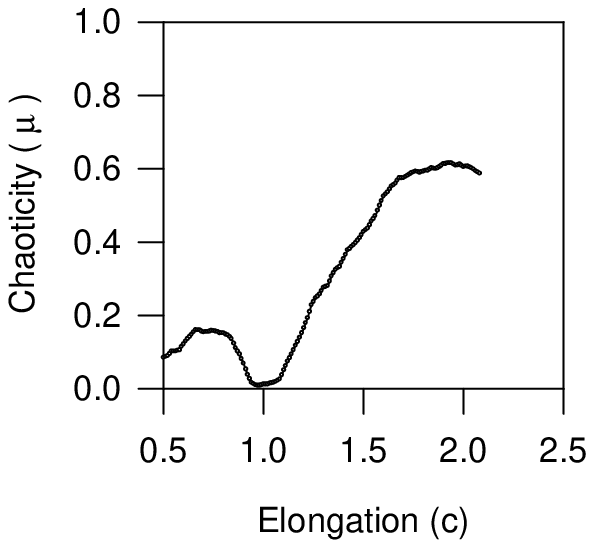}}
\caption
{\label{fig1}Variation of chaoticity with elongation.}
\end{figure}

We have calculated the chaoticity for a range of shapes from oblate to the
scission configuration (at $c=2.08$ where the neck radius becomes zero) at
small steps of $c$, the elongation coordinate. Figure \ref{fig1}
shows the calculated
values of chaoticity which will be subsequently employed to obtain the
chaos weighted wall formula friction. It is important to note here that
chaoticity is very small for near spherical shapes ($c \sim 1$). This
immediately implies, through Eq.\ref{seqn1},  a strong suppression
of the original
wall formula
friction for compact shapes of the compound nucleus. Chaoticity however
increases as the shape becomes more oblate or changes towards scission
configuration. We find here that the full chaotic regime ($\mu =1 $)
in single particle dynamics is not
reached even at the scission configuration. This observation is specific to the
shape parametrization (Eq.\ref{seqn2}) used in the present calculation. It was
observed earlier that the value of chaoticity reaches $1$ near the scission
point when Legendre polynomial $P_{2}$ deformed quadrupole shapes were
considered \cite{pal3}. It is difficult to speculate at this point to what
extent the final calculated observables will be sensitive to the choice of
the shape parametrisation.

\begin{figure}[htb]
\centering{\
\epsfig{figure=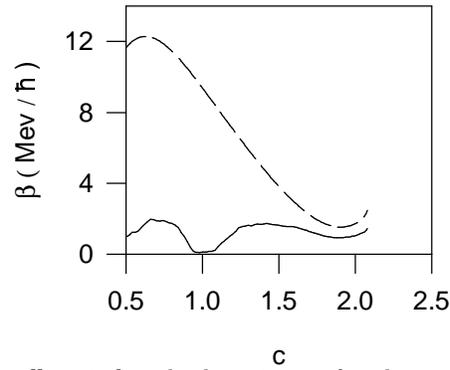}}
\caption
{\label{fig2}Variation of reduced friction coefficient $\beta$
with elongation $c$
for chaos weighted wall formula (full line) and wall formula (dashed line)
frictions.}
\end{figure}

We shall use the following expression to calculate the wall formula friction
coefficient \cite{sierk2},
\begin{eqnarray}
\eta_{wf}(c)&=&{1 \over 2} \pi \rho_m {\bar v} \left\{\int_{z_{min}}^{z_N}
{\left(
\frac{\partial \rho^2}{\partial c} + \frac{\partial \rho^2}{\partial z}
\frac{\partial D_1}{\partial c}\right)}^2 {\left[\rho^2
+ {\left({1 \over 2}\frac{\partial
\rho^2}{\partial z}\right)}^2\right]}^{-{1 \over 2}}dz\right .\nonumber\\
&&+\left .\int_{z_N}^{z_{max}} {\left(
\frac{\partial \rho^2}{\partial c} + \frac{\partial \rho^2}{\partial z}
\frac{\partial D_2}{\partial c}\right)}^2 {\left[\rho^2 + {\left({1 \over 2}
\frac{\partial
\rho^2}{\partial z}\right)}^2\right]}^{-{1 \over 2}}dz \right\},
\label{seqn7}
\end{eqnarray}
where $\rho_{m}$ is the mass density of the nucleus, $\bar{v}$ is the average
nucleon speed inside the nucleus and $D_{1}$, $D_{2}$ are the positions of the
centers of mass of the two parts of the fissioning system relative to the
center of mass of the whole system. $z_{min}$ and $z_{max}$ are the two
extreme ends of the nuclear shape along the $z$ axis and $z_{N}$ is the
position of the neck plane which divides the nucleus into two parts.
The chaos weighted wall formula friction is subsequently obtained from
Eq.\ref{seqn1} as $\eta_{cwwf} (c) = \mu (c) \eta_{wf} (c) $.
Defining a quantity $\beta (c)= \eta (c)/ m(c)$ as the reduced friction
coefficient, its dependence on the elongation coordinate is shown in
Fig.\ref{fig2}
for both WF and CWWF frictions for the $^{200}Pb$ nucleus. The reduction in
the strength of the wall friction due to chaos considerations is evident
from this figure.

\begin{figure}[htb]
\centering{\
\epsfig{figure=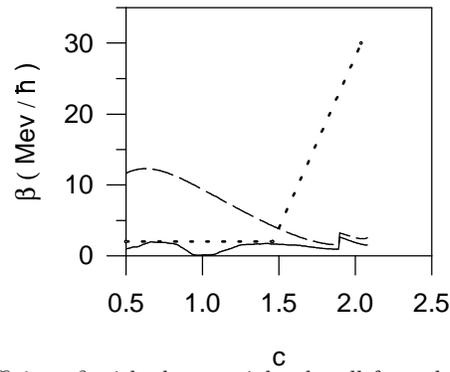}}
\caption
{\label{fig3}Reduced one-body friction coefficient $\beta$
with chaos weighted wall
formula (full line) and wall formula (dashed line) frictions. The
phenomenological reduced coefficient (dotted line) from
Ref.[9] is
also shown.}
\end{figure}

We shall now consider the role of window friction in one-body dissipation.
The window friction is expected to be effective after a neck is formed in
the nuclear system \cite{sierk2}.
Further, the radius of the neck connecting the two  future fragments
should be sufficiently narrow
in order to enable a particle which has crossed the window
from one side to the other to remain within the other fragment
for a sufficiently
long time.
This is necessary to allow the particle to suffer
enough collisions within the other side and make
the energy transfer irreversible.
It therefore appears that the window friction should be very nominal when
neck formation just begins. Its strength should however increase as the
neck becomes narrower reaching its classical value when the neck radius
becomes much smaller than the typical radii of the fragments.
Little is however known regarding the detailed nature of
such a transition. We shall therefore refrain from making any further
assumption regarding the onset of window friction. Instead,
we shall define a critical
elongation coordinate $c_{win}$ beyond which the window friction will be
switched on. The window friction coefficient will then be given as,
\begin{equation}
\eta_{win}(c) = \theta(c - c_{win}) {1 \over 2} \rho_m {\bar v}
{\left(\frac{\partial R}{\partial c}\right)}^2 \Delta \sigma,
\label{seqn8}
\end{equation}
where
\begin{eqnarray}
\theta(c - c_{win}) &=& 0\quad for \quad c < c_{win},\nonumber\\
                    &=& 1\quad for \quad c \ge c_{win},\nonumber
\end{eqnarray}
and $R$ is the distance between centers of mass of future fragments  and
$\Delta \sigma$ is the area of the window between the two parts of the
system. The full one-body friction will now be written as,
\begin{equation}
\eta(c)= \eta_{wall} (c) + \eta_{win} (c),
\label{seqn9}
\end{equation}

\noindent and in what follows, we will use either $\eta_{wf}$ or
$\eta_{cwwf}$ for
$\eta_{wall}$ in the above expression.
For the window friction, the value of $c_{win}$
is taken as 1.9 where the neck radius is half of the fragment radius.
Figure \ref{fig3} shows the reduced one-body
friction coefficients.
The phenomenological reduced friction obtained in Ref.\cite{fro3} is
also shown in Fig.\ref{fig3}. Though the one-body friction with CWWF agrees
qualitatively with the phenomenological friction for $c<1.5$, it is beyond its
scope to
explain the steep increase of phenomenological friction for $c>1.5$.
It may be noted however that the compulsion of having a very strong friction at
large deformations was to allow enough neutrons to evaporate during the
saddle to scission transition ( i.e. after fission has taken place)
in order to fit the experimental prescission neutron
multiplicities for very heavy nuclei \cite{fro3}. Therefore, the role of a
very strong
friction beyond the saddle point will not be significant for fission rates
which is of our present concern.

\section {RESULTS}

 With all the necessary inputs defined as above, the Langevin
 equation (Eq.\ref{seqn3})
is numerically integrated following the procedure outlined in Ref.\cite{abe}.
A very small time step of $0.005 \hbar /MeV$ for numerical integration
is used in the present work.
The numerical stability of the results is checked by repeating a few
calculations with still smaller time steps. The initial distribution
of the coordinates and momenta are assumed to be
close to equilibrium and hence the initial values of $(c,p)$ are
chosen from sampling random numbers following the Maxwell-Boltzmann
distribution. Starting with a given total
excitation energy ($E^{*}$) and angular momentum ($l$) of the compound nucleus,
the energy conservation in the following form,
\begin{equation}
E^{*}=E_{int}+V(c)+p^{2}/2m
\label{seqn10}
\end{equation}
 gives
 the intrinsic excitation energy $E_{int}$ and the corresponding nuclear
 temperature $T=(E_{int}/a)^{1/2}$ at each integration step.
 The centrifugal potential is included in $V(c)$ in the above equation.
 A Langevin
 trajectory will be considered as undergone fission if it reaches the
 scission point ($c_{sci}$) in course of its time evolution. The calculations
 are repeated for a large number (typically 100,000 or more) of trajectories
 and
 the number of fission events are recorded as a function of time. From these,
 the fission rates can be easily evaluated \cite{ab2}.

\begin{figure}[htb]
\centering{\
\epsfig{figure=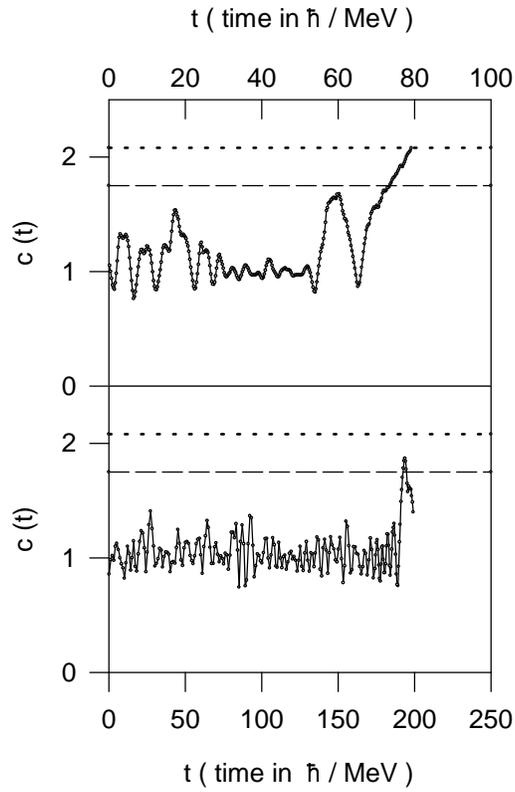}}
\caption
{\label{fig4}The upper panel shows a typical Langevin trajectory
reaching the scission
point (dotted line). The lower panel shows a trajectory which returns to
the potential pocket after crossing the saddle point (dashed line).}
\end{figure}

\begin{figure}[htb]
\centering{\
\epsfig{figure=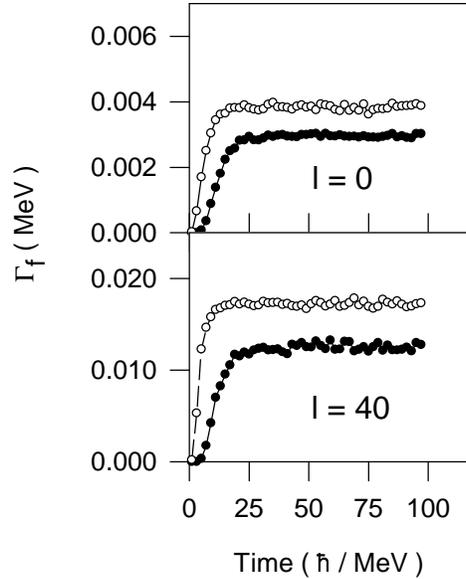}}
\caption
{\label{fig5}Time development of fission widths for compound
nuclear spins of
0 and 40 (in $\hbar$ unit). Open circles correspond to trajectories for
which saddle point crossing is considered as fission. Filled circles
represent trajectories which reach the scission point.}
\end{figure}

 A typical Langevin trajectory which has reached the scission point and has
 ended up as a fission event is
 shown in Fig.\ref{fig4} (upper panel). Another trajectory, the kind of
 which is
 less frequent, is shown in the lower panel of the same figure. The Langevin
 trajectory in this case crosses the saddle point and after spending some time
 beyond the saddle point drifts back into the potential pocket again. Such
 trajectories may or may not finally reach the scission point within the
 observation time and corresponds to a to-and-fro motion
 across the saddle and
 essentially portrays the stochastic nature of the dynamics.
 This point is further illustrated in Fig.\ref{fig5} where
 time development of the
 fission rates are plotted. Two different criteria are used to define a fission
 event here. The filled circles correspond to fission events defined as those
 trajectories reaching the scission point whereas the open circles correspond
 to those crossing the saddle point. The fission rate is very small for both
 the cases at the beginning when the compound nucleus is just formed and
 the Langevin dynamics
 has just been turned on. Subsequently the fission rate grows with time and
 after a certain equilibration time it reaches a stationary value which
 corresponds to a steady flow across the barrier.
 The fission rate defined at the saddle point reaches the stationary value
 earlier than that defined at the scission point. The time difference
 between them gives the average time of descent from the saddle to
 the scission. This observation was also
 made in earlier works \cite{wada}. The main purpose of the present
 discussion is to investigate the role of backstreaming in the fission
 process. It is observed in Fig.\ref{fig5} that the stationary
 fission rate at
 saddle point is higher than that at the scission point. The difference between
 these two stationary rates can be regarded as due to backstreaming.
 The backstreaming is thus small compared to the steady outward flow though
 it is not negligible. This also shows that crossing the saddle point is
 not an adequate criteria for fission in stochastic calculations and can lead
 to an overestimation of the fission rate.

\begin{figure}[htb]
\centering{\
\epsfig{figure=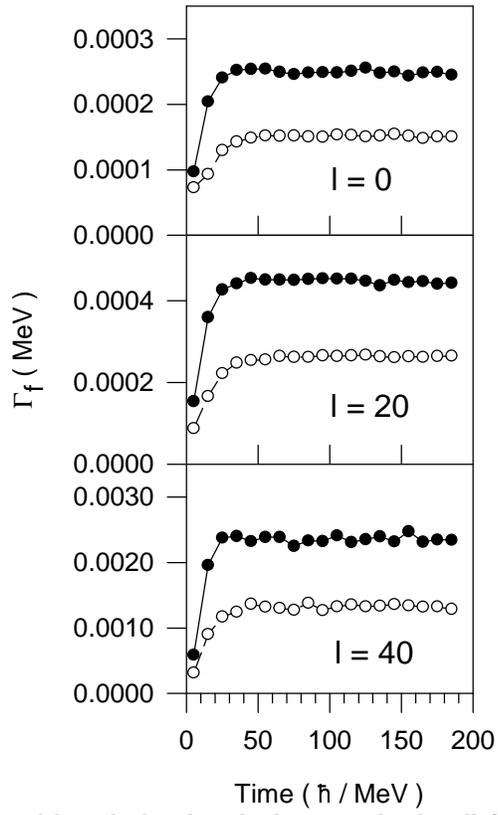}}
\caption
{\label{fig6}Time development of fission widths calculated with
chaos weighted wall formula
(filled circles) and wall formula (open circles) frictions for different
compound nuclear spins $l$ (in $\hbar$ unit).}
\end{figure}

 We shall now compare fission rates calculated with chaos weighted wall-and-
 window friction with those obtained with wall-and-window friction
 (Eq.\ref{seqn9}).
 Figure \ref{fig6} shows the fission widths at three spins
 of the compound nucleus
 $^{200}Pb$. The effect of suppression in the chaos weighted wall formula
 shows up as an enhancement by about a factor of 2 of the stationary
 fission rates. Similar
 enhancement of the stationary fission rate calculated with chaos weighted
 wall-and-window friction in comparison with that obtained with wall-and-window
 friction are also observed for a wide range of compound nuclear spin
 and temperature. The enhancement factor (of about 2) remains almost the
 same when different choices of $c_{win}$ are used in the window
 friction (Eq.\ref{seqn8}).

\begin{figure}[htb]
\centering{\
\epsfig{figure=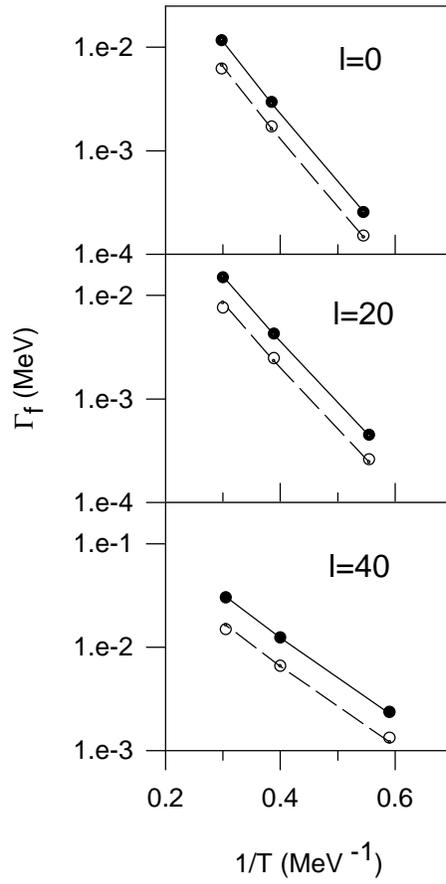}}
\caption
{\label{fig7}Temperature dependence of stationary fission widths
calculated with chaos weighted wall formula
(filled circles) and wall formula (open circles) frictions for different
compound nuclear spins $l$ (in $\hbar$ unit). The lines are fitted as
explained in the text.}
\end{figure}

\begin{figure}[htb]
\centering{\
\epsfig{figure=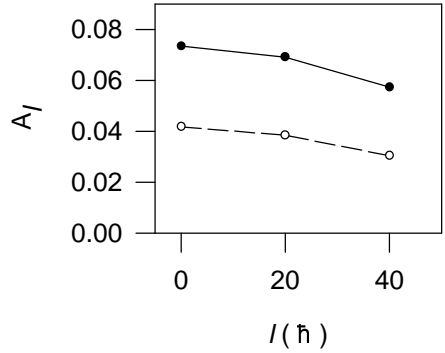}}
\caption
{\label{fig8}Variation of the parameter $A_{l}$ with compound
nuclear spin $l$.}
\end{figure}

 We next systematically extracted the stationary fission widths at different
 temperatures for a given spin of the compound nucleus. This was done by
 taking the average of the fission rates in the plateau region.
 These fission rates are essentially the Kramers' limit of the Langevin
 equation under consideration and we expect the stationary fission
 widths $\Gamma_{f}$ to depend upon the
 temperature  $T$ as $\Gamma_{f}(l,T)= A_{l} \exp (-b_{f}/T)$ for a given
 spin ($l$)
 of the compound nucleus where $b_{f}$ is the height of the fission barrier
 in the free energy profile and $A_{l}$ is a parameter. Such a dependence
 of stationary fission widths on temperature
 was indeed found and is shown in Fig.\ref{fig7}.
 The parameter $A_{l}$ can now be extracted by fitting the calculated fission
 widths with the above expression.
  Subsequently we looked into
 the dependence of the parameter $A_{l}$ on $l$, a few typical plots of which
 are shown in Fig.\ref{fig8}.
 Using these values of $A_{l}$, one can now obtain this parameter value
 for any arbitrary spin by interpolation. Even with a limited number of
 calculated values, the interpolated values will be quite reliable since
 $A_{l}$ depends on $l$ rather weakly as can be seen in Fig.\ref{fig8}.
 Consequently
 it becomes possible to extract the fission width of a compound nucleus of
 any given temperature and spin from a set of a limited number of
 calculated widths. This fact will be very useful in statistical model
 calculations where fission widths are required at numerous values of
 temperature and spin which are encountered during evolution of
 a compound nucleus.
 Therefore in such cases where analytical expressions
 for fission widths cannot be obtained, the above systematics can generate
 fission widths from a limited set of calculations.

\begin{figure}[htb]
\centering{\
\epsfig{figure=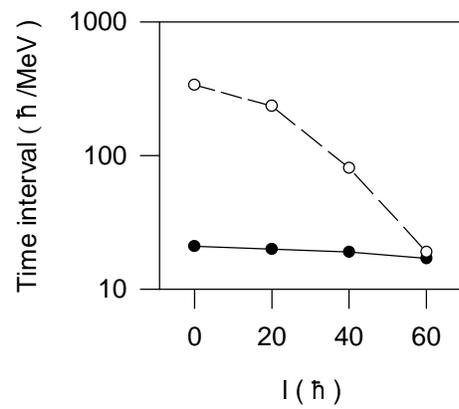}}
\caption
{\label{fig9}Dependence of equilibration time $\tau_{eq}$ (filled circles)
and
fission life time $\tau_{f}$ (open circles) on compound nuclear spin $l$.}
\end{figure}

 Two time scales are of physical significance in the Langevin description
 of dynamics of fission. One is the equilibration time $\tau_{eq}$, the time
 required to attain a steady flow across the barrier. The other is the
 fission life time $\tau_{f}= \hbar / \Gamma_{f}$.
 Figure \ref{fig9} shows these
 time intervals for different values of spin of the compound nucleus
 $^{200}Pb$. At very small values of spin, the fission life time is many times
 longer than the equilibration time. This means that a statistical theory
 for compound nuclear decay is applicable in such cases. On the other hand,
 $\tau_{eq}$ and $\tau_{f}$ become comparable at  higher values of the
 compound nuclear spin and this corresponds to a dynamics dominated
 decay of the compound nucleus. Statistical models are not meaningful
 in these cases and dynamical descriptions such as Langevin equation
 become essential for fission of a compound nucleus.

 \section{SUMMARY AND OUTLOOK}

 In the preceeding sections, we have presented a systematic study of fission
 dynamics using Langevin equation. Among the various physical inputs
 required for solving the Langevin equation, we paid particular attention
 to the dissipative force for which we chose the wall-and-window
 one-body friction. We
 used a modified form of wall friction, the chaos weighted wall formula, in
 our calculation. The chaos weighted wall formula took into account
 the nonintegrabilty of
 single particle motion in the nucleus and it resulted in a strong suppression
 of friction strength for near spherical shapes of the nucleus. The fission
 widths calculated with chaos weighted wall formula turned out to be about
 twice the widths calculated with the normal wall formula friction. The
 chaos weighted wall friction thus enhances the fission rate substantially
 compared to that obtained with normal wall friction.

 We further made a parametric representation of the calculated fission widths
 in terms of the temperature and spin of the compound nucleus. It was found
 that this parametrised form can be well determined from the
 fission widths calculated over a grid of spin and temperature values
 of limited size.
 This fact would make it possible to perform statistical model
 calculation of the decay of a highly excited compound nucleus where the
 fission
 widths are to be determined from a dynamical model such as the Langevin
 equation. When the friction form factor has a strong shape dependence as in
 the chaos weighted wall formula, the corresponding fission widths cannot be
 obtained in an analytic form. In such cases, the frequently required values
 of the fission width in a statistical model calculation can be made
 economically accessible through the parametrised representation of the
 fission width which has to be obtained in a separate calculation similar to
 the present one. We shall report on such applications of the parametrised
 fission widths in compound nuclear decay in our future works.

\eject

\end{document}